# Variable-length Hill Cipher with MDS Key Matrix

**Kondwani Magamba [†], Solomon Kadaleka [††] and Ansley Kasambara [††]**
magambakho@yahoo.co.uk    skadaleka@poly.ac.mw    akasambara@poly.ac.mw
[†]Department of Mathematics, Catholic University of Malawi, Montfort Campus, Malawi
[††]Department of Mathematics and Statistics, Polytechnic College, University of Malawi, Blantyre, Malawi

**Summary**
The Hill Cipher is a classical symmetric cipher which breaks plaintext into blocks of size $m$ and then multiplies each block by an $m \times m$ key matrix to yield ciphertext. However, it is well known that the Hill cipher succumbs to cryptanalysis relatively easily. As a result, there have been efforts to strengthen the cipher through the use of various techniques e.g. permuting rows and columns of the key matrix to encrypt each plaintext vector with a new key matrix. In this paper, we strengthen the security of the Hill cipher against a known-plaintext attack by encrypting each plaintext matrix by a variable-length key matrix obtained from a Maximum Distance Separable (MDS) master key matrix.

## 1. Introduction

The Hill cipher is a classical symmetric cipher based on matrix transformation. It has several advantages including its resistance to frequency analysis and simplicity due to the fact that it uses matrix multiplication and inversion for encryption and decryption. However, it succumbs to the known-plaintext attack [4] and as such there have been efforts to strengthen the cipher through the use of various techniques which have improved the security of the cipher quite significantly [1],[2],[3]. Unlike the original Hill cipher the proposed modifications of the Hill cipher have practical applications including image encryption see for example [6].

The method employed in this paper aims at generating dynamic variable-length key matrices from a given shared MDS master key matrix. With the generated matrices, each plaintext matrix is encrypted using a different key matrix and this renders the ciphertext immune to a known-plaintext attack.

The rest of the paper is divided as follows; in Section 2 the basic concept of a Hill cipher is outlined, a literature review is in Section 3, Sections 4 and 5 present a discussion on MDS matrices and in Section 6 the proposed algorithm is described.

## 2. Hill Cipher

Let us consider a plaintext string $P$ of length $l$ defined over an alphabet of order $n$. Each letter is mapped to an element of $\mathbb{Z}_n$. Often $n = 26$ is used and the correspondence mostly used is: $A = 0, B = 1, \ldots, Z = 25$. Additionally, an $m \times m$ matrix $K$ with entries from $\mathbb{Z}_n$ is chosen as the secret key. The encryption and decryption is then given as;

$$C = KP \bmod n \qquad P = KC \bmod n \qquad (1)$$

where $C$ is the ciphertext. For encryption to be possible $P$ is divided into substrings of length $m$. If $l$, the length of the plaintext, is not divisible by $m$ then the plaintext block must be padded with extra characters. Another complication arises when $K$ is not invertible. A necessary and sufficient condition for a matrix $K$ over $\mathbb{Z}_n$ to be invertible is that the determinant of $K$ be non-zero and co-prime to $n$. See [7] for the size of the keyspace of a Hill cipher.

## 3. Literature Review

Several research works have been done with an aim to improve the security of Hill cipher. For example, a method HCM-PT found in [5] tries to make Hill cipher secure by using dynamic key matrix $K_t$ obtained by random permutations $M_t$ of columns and rows of the master key matrix and transfers an encrypted plaintext and encrypted permutation vector $\underline{u} = Kt$ to the receiving side. The number of dynamic keys generated is $m!$ where $m$ refers to the size of the key matrix.

HCM-PT   Encryption
$$K_t = M_t K M_t^{-1}$$
$$C = K_t P$$
$$\underline{u} = Kt$$

Even though this method thwarts the known-plaintext attack on the plaintext, it is vulnerable to known-plaintext attack on the permutation vector. This prompted [11] to modify HCM-PT.

The modification to HCM-PT called SHC-M found in [11] works the same way as HCM-PT but does not transfer the permutation vector, instead both the sender and the receiver use a pseudo-random permutation generator, $t = PRPG(seed, n)$ and only the number of the necessary permutation is transferred to the receiver. The number of dynamic keys is the same as HCM-PT.

```
             SHC-M Encryption
1. Sender, S, selects a number, n
2. calculates t = PRPG(seed, n)
3. S → R: C, n xor SEED
4. Proceed as in HCM − PT
```

Another modification to the Hill cipher is found in [6] which tries to improve the security of Hill cipher by using a one-time-one key matrix to encrypt each plaintext block. This unique key is computed by multiplying the current key with a secret Initial Vector (IV). However, [10] shows that the method is prone to a known-plain text attack.

The method presented here works the same way as SHC-M but our method requires that the master key matrix be MDS. With an MDS master key, the key space is bigger than that of SHC-M and our method can be used to encrypt plaintext blocks of variable length. As far as we know this is the first instance of a variable length Hill cipher.

## 4. MDS Matrices

Let $C$ be a linear code of length $n$, dimension $k$, and distance $d$ with parameters $[n, k, d]$. A generator matrix $G$ for a linear $C$ is a $k \times n$ matrix whose rows form a basis for $C$. Linear codes obey the Singleton bound, $d \leq n - k + 1$. If a linear code meets the Singleton bound, $d = n - k + 1$, then it is called a Maximum Distance Separable code or MDS code. Also, an $[n, k, d]$-error correcting code with generator matrix $G = [I_{k \times k} | A]$ where $I_{k \times k}$ is the $k \times k$ identity matrix, and $A$ is a $k \times (n - k)$ matrix, is MDS if and only if every square submatrix of $A$ is nonsingular [8]. In most applications e.g. [12] the code $C$ is taken as a linear code with parameters $[2k, k, k + 1]$ and generator matrix $G = [I_{k \times k} | A_{k \times k}]$. When we take a generator matrix of a linear MDS code with parameters $[2k, k, k + 1]$ as $G = [I_{k \times k} | A_{k \times k}]$, the $k \times k$ matrix $A$ is the matrix we call MDS matrix. MDS matrices are an important building block adopted by different block ciphers as they guarantee fast and effective diffusion in a small number of rounds. See for example [12]. In this paper MDS matrices are used as key matrices for a Hill cipher. This modification increases the key space of SHC-M because in an MDS matrix every square submatrix is non-singular which is the requirement for a key matrix of a Hill cipher.

## 5. Generating MDS Matrices

There are different methods of generating MDS matrices e.g. using Cauchy and Hadamard matrices. MDS matrices can also be generated exhaustively but probably the most popular method is to use Reed Solomon Codes.

## 6. Variable-Length Hill Cipher

As pointed out earlier SHC-M has a dynamic key space (NDK) of $m!$ but in this section we show that by requiring that the key matrix be MDS we increase the key space of SHC-M. We also develop a way of encrypting variable-length plaintext blocks.

6.1 Enlarging the Key space

We will work as follows; suppose the secret shared MDS matrix $K$ is of size $m \times m$ where $m \geq 3$. The cases $m = 1, 2$ are trivial. Then for each of the $m!$ permutations of the columns of $K$ we can form $j \times j$ invertible submatrices where $j \in [2..m]$. Each of these matrices can be used for encryption in our method. This can be done by choosing columns of a required length from any of the permutations.

Alternatively, we can start by choosing $j$ columns from a total of $m$ columns. We then form $j \times j$ invertible submatrices from the $\binom{m}{j}$ columns. Each possible submatrix is then permuted to give $j!$ permuted submatrices. We do this for all possible matrix lengths $j \in [2..m]$. It is not difficult to see that the total number of dynamic matrices (NDK) is;

$$\text{NDK} = \sum_{j=2}^{m} \binom{m}{j}^2 j!$$

Obviously, this number gets bigger as $m$ increases. AS fra as we know, the number of dynamic keys found here is the best of all the known modifications of the Hill cipher. The best thing with this method is that it applies to any Hill cipher modification which proposes dynamic matrix keys.

We illustrate the method with an example. Let

$$K = \begin{pmatrix} k_{11} & k_{12} & k_{13} \\ k_{21} & k_{22} & k_{23} \\ k_{31} & k_{32} & k_{33} \end{pmatrix}$$

be a $3 \times 3$ MDS secret key matrix. Then applying the formula above we have 18 matrices of order $2 \times 2$ and 6 matrices of order $3 \times 3$ giving a total of 24 dynamic matrices. We now show how to generate all the 24 matrices.

We begin by choosing 2 columns from $K$. We get the following matrices, $\begin{pmatrix} k_{11} & k_{12} \\ k_{21} & k_{22} \\ k_{31} & k_{32} \end{pmatrix}, \begin{pmatrix} k_{11} & k_{13} \\ k_{21} & k_{23} \\ k_{31} & k_{33} \end{pmatrix}$ and $\begin{pmatrix} k_{12} & k_{13} \\ k_{22} & k_{23} \\ k_{32} & k_{33} \end{pmatrix}$. From these matrices we form $2 \times 2$ matrices e.g from $\begin{pmatrix} k_{11} & k_{12} \\ k_{21} & k_{22} \\ k_{31} & k_{32} \end{pmatrix}$ we get $\begin{pmatrix} k_{11} & k_{12} \\ k_{21} & k_{22} \end{pmatrix}, \begin{pmatrix} k_{11} & k_{12} \\ k_{31} & k_{32} \end{pmatrix}$ and $\begin{pmatrix} k_{21} & k_{22} \\ k_{31} & k_{32} \end{pmatrix}$ and column-permuted versions of these matrices. Continuing in the same fashion and considering the 6 permutations of $K$ we get all 24 dynamic matrices.

It can be seen that the key space of SHC-M is now $\sum_{j=2}^{m}\binom{m}{j}^2 j!$ which is a significant increase from $m!$. This number doubles when we permute the dynamic matrices in a row wise manner. The key space can also be increased by using $j \times j$ matrices to form $2j \times 2j$ matrices for encryption. For example, if we have the 4 $j \times j$ matrices $K_1, K_2, K_3$ and $K_4$ we may, $wlog$, form a new matrix;

$$K_* = \begin{pmatrix} K_1 & K_2 \\ K_3 & K_4 \end{pmatrix}$$

6.2. Encryption and Decryption

Encryption is the same as SHC-M but we require that we choose a control key (preferably binary) to govern encryption, see [13]. This key determines the length of block to encrypt i.e. when to encrypt a $2 \times 2$ block, a $3 \times 3$ block and so on. To avoid having too many secret keys we can use the seed as the control key but a problem may arise when we have to expand the seed due to a long plaintext. Once we have the binary control key we set up a correspondence between binary strings and block lengths. For example, if $m = 4$ then the possible lengths of subkeys are 2, 3 and 4 therefore we can have the following set up;

$$01 \rightarrow 2, 10 \rightarrow 3 \text{ and } 11 \rightarrow 4.$$

This means we read the control key two bits at a time and encrypt block lengths accordingly. It is therefore very important that the control key be random to prevent attacks. We now give a full description of the proposed method:

---
**Algorithm 1:** Encryption
**Data:** Plaintext P, K, SEED, Control Key
**Result:** Cipher Text, C
**begin**
$\quad r \leftarrow$ block number
$\quad t = \text{PRPG(SEED, } r) \leftarrow$ a permutation order
$\quad M_t \leftarrow$ a permutation from $t$
$\quad K_t = M_t K M_t^{-1} \leftarrow$ encryption matrix
$\quad$ Determine block length $m$ from control key
$\quad$ Obtain a length $m$ encryption matrix $K_m$ from $K_t$
$\quad$ Get plaintext of length $m$, $P_m$, from P
$\quad$ Encrypt using $C = K_m P_m$
$\quad$ Send C, r xor SEED
**end**

---
**Algorithm 2:** Decryption
**Data:** Ciphertext C, K, SEED, Control Key
**Result:** Plaintext, P
**begin**
$\quad$ Calculate $t$ from SEED and $r$
$\quad$ Calculate $M_t$; Calculate $K_t = M_t K M_t^{-1}$
$\quad$ Get $m$ from control key
$\quad$ Obtain $K_m$ from $K_t$
$\quad$ Decrypt using $P_m = K_m^{-1} C_m$
**end**

---

From the encryption algorithm one easily sees that the sender and receiver need to find a way of obtaining a length $m$ encryption matrix $K_m$ from $K_t$. One way could be to take matrices in a left-right then top-bottom manner. It is worth noting that there are different ways of choosing variable-length matrices from $K_t$ and this strengthens the proposed algorithm even more.

## 7. Security Analysis

The main limitation with Hill Cipher is that it is prone to a known plaintext attack, that is, if an attacker has $m$ distinct plaintext and ciphertext pairs then they can retrieve the key by solving $K = P^{-1}C$ or $K = CP^{-1}$ depending on the encryption equation used [6]. If $P$ is not invertible, then other sets of $m$ plaintext-ciphertext pairs have to be tried. By encrypting variable-length blocks of plaintext the proposed method is safe from known plaintext attacks. Also the proposed method inherits immunity from a cipher-text only attack from the original Hill cipher. The proposed method does not, however, introduce sufficient confusion and diffusion to be ranked among the best and efficient symmetric key cryptosystems. It is worth pointing out that the proposed algorithm relies on many matrix transformations and this slows down the algorithm.

## 8. Conclusion

We have presented a variable-length Hill cipher which is a modification of the original Hill cipher. It is also based on SHC-M which is itself also a modification of the Hill cipher. Our method strengthens the Hill cipher against known-plaintext and ciphertext only attacks. The former is due to the use of dynamically changing key matrices. Our method is better than the methods found in the review in that it has a larger key space. On the downside our method is not efficient due to the use of many matrix operations.